\documentclass[twocolumn,aps,floatfix]{revtex4}
\usepackage{amssymb}
\usepackage{graphicx}

\begin{document}

\title{Thermally induced instability of a doubly quantized vortex in a 
       Bose-Einstein condensate}

\author{Krzysztof Gawryluk,$\,^1$ Miros{\l}aw Brewczyk,$\,^1$ and 
        Kazimierz Rz{\c a}\.zewski$\,^2$}

\affiliation{\mbox{$^1$ Instytut Fizyki Teoretycznej, Uniwersytet w Bia{\l}ymstoku,
                        ulica Lipowa 41, 15-424 Bia{\l}ystok, Poland}  \\
\mbox{$^2$ Centrum Fizyki Teoretycznej PAN, Aleja Lotnik\'ow 32/46, 02-668 Warsaw,
           Poland}   }

\date{\today}

\begin{abstract}
We study the instability of a doubly quantized vortex topologically imprinted 
on $^{23}$Na condensate, as reported in recent experiment 
[Phys. Rev. Lett. \textbf{93}, 160406 (2004)]. We have performed numerical
simulations using three-dimensional Gross-Pitaevskii equation with classical 
thermal noise. Splitting of a doubly quantized vortex turns out to be a process
that is very sensitive to the presence of thermal atoms. We observe that even very 
small thermal fluctuations, corresponding to $10$ to $15\%$ of thermal atoms, cause 
the decay of doubly quantized vortex into two singly quantized vortices in tens 
of milliseconds. As in the experiment, the lifetime of doubly quantized vortex is 
a monotonic function of the interaction strength.
\end{abstract}

\maketitle

Experimental studies of vortices in Bose-Einstein condensates have revealed 
their peculiar properties related to the quantized circulation \cite{vortex},
originally predicted by Onsager and Feynman in the context of rotating
superfluid $^4$He \cite{circ} and further explored in Refs. \cite{others1} 
and \cite{others2}. The quantization of the circulation is a 
manifestation of the existence of a macroscopic wave function. Direct 
$2\pi$-change of the phase of the condensate wave function when going around 
the vortex core was experimentally demonstrated by using the interferometric 
technique \cite{Dalibard}. The quantized vortex can not just disappear, it can 
leave the condensate or annihilate with a vortex having the opposite circulation. 
There exists another route for vortices that have multiple topological charge, 
they can also split into several vortices having smaller charges decreasing in 
such a way the energy of the system.

In recent experiments \cite{Ketterle1,Ketterle2} doubly quantized vortices
have been imprinted in Bose-Einstein condensate of $^{23}$Na atoms. A novel
approach allowing to create vortices with multiple topological charge was
implemented \cite{Isoshima}. In this new method, as opposed to dynamical
phase-imprinting techniques like rotating the atomic cloud in an anisotropic
trap or stirring the condensate with the help of a laser beam, vortices are
generated by adiabatically reversing the magnetic bias field along the
trap axis. This topological phase imprinting technique leads to vortices
displaying winding numbers $2$ or $4$ depending on the hyperfine state
the sodium condensate was prepared in ($|1,-1>$ and $|2,+2>$, respectively)
\cite{Ketterle1}. In Ref. \cite{Ketterle2} the authors investigate the
evolution of doubly quantized vortices by using a tomographic imaging
method \cite{tomographic}. They observe the decay of doubly quantized vortex 
into singly quantized vortices and suggest the possible explanation of this 
splitting as being a result of dynamical instability, however, not specifying
the character of the perturbation seeding the instability.

Existing theoretical explanation of decay of doubly quantized vortices
involves the analysis of stability of the vortex at zero temperature in 
terms of eigenmode spectrum of the Bogoliubov equations in two- and 
three-dimensional geometry as well as the numerical solution of the 
Gross-Pitaevskii equation in slightly anisotropic trap \cite{Mottonen}. 
Two-dimensional Bogoliubov eigenvalue spectrum shows complex frequencies 
for certain values of the interaction strength \cite{Mottonen, Eberly}. 
In fact, quasiperiodic behavior is found -- the stability windows are followed 
by the instability regions. Neither the anisotropy of the trap nor the 
phenomenologically introduced dissipation is able to force the decay of doubly 
quantized vortex when there are no complex eigenvalues ({\it i.e.}, the stability 
window conditions are fulfilled) \cite{Mottonen}. However, in the experiment the 
monotonic increase of the lifetime of the vortex with the strength of the 
nonlinearity is observed \cite{Ketterle2}. It turns out that such behavior is 
attributed to the three-dimensional geometry of the system and its origin could
be due to the presence of trap anisotropy (as suggested in Ref. \cite{Mottonen}) 
or the thermal noise (as claimed by this Letter).

In this Letter we show that only thermal fluctuations, not other disturbances 
(as, for example, due to the confinement anisotropy), lead to the decay times 
comparable with that reported in the experiment \cite{Ketterle2}. To this end, we 
have performed numerical simulations using the three-dimensional Gross-Pitaevskii 
equation in the version described in Ref. \cite{Warsaw}, {\it i.e.}, with a 
classical thermal noise. We find that even extremely small thermal fluctuations 
dramatically accelerate the decay of doubly quantized vortex into two singly 
quantized vortices. Increasing the number of thermal (uncondensed) atoms already 
to $10$ to $15\%$ reduces the lifetime of the vortex below $100\,$ms. Therefore, we  
argue that although the authors of Ref. \cite{Ketterle2} report that the experiment 
is performed under the condition of no discernible thermal atoms presence, the decay 
of the vortex is triggered by thermal rather than quantum fluctuations.

We describe the system of degenerate bosonic atoms in terms of the classical field 
that is the complex function representing all atoms not only those within the 
condensed faction. At zero temperature all atoms are condensed and the classical 
field becomes the condensate wave function which satisfies the Gross-Pitaevskii 
equation. However, as it was argued in Ref. \cite{Warsaw}, the same equation is 
fulfilled by the classical field at finite temperatures, although in this case it 
must be interpreted in a different way. The high energy solution of the Gross-Pitaevskii 
equation describes both the condensed and thermal atoms. To get correct physical
interpretation of the classical field one has to perform averaging over time or 
space of corresponding single-particle density matrix. The condensate wave function
is just the dominantly populated eigenmode of the averaged single-particle density 
matrix. Other modes represent thermal atoms.

We start our simulations with the wave function of the cigar-shaped condensate with 
a vortex imprinted along the symmetry axis. The topological charge of the vortex equals 
$2$. The trap parameters are the same as in the experiment of Ref. \cite{Ketterle2},
{\it i.e.}, the radial (axial) trap frequency equals $220$\,Hz ($3$\,Hz). The number of 
atoms is changed in such a way that the values of the control parameter defined as $a n_z$, 
where $a$ is the scattering length and $n_z=\int |\,\psi(x,y,z=0)\,|^2 dx\, dy$ is the 
linear atom density along the symmetry axis taken at the center of the trap, varies up 
to about $14$ (see Ref. \cite{Ketterle2}). To find the condensate wave function with the 
vortex we solve the Gross-Pitaevskii equation in imaginary time. The numerical code 
implements an operator splitting technique as described in \cite{recipes} and works 
on two-dimensional grid since the phase of the vortex state, $e^{i 2\phi}$, can be 
separated off analytically.

Next, we randomly disturb the condensate wave function, {\it i.e.}, introduce the 
thermal noise into the classical field. Since the classical field is a complex function 
we perturbed independently the amplitude and the phase. Both approaches give very similar 
results. The evolution of the classical field $\psi$ is governed by the time-dependent 
Gross-Pitaevskii equation (see Ref. \cite{Warsaw})
\begin{equation}
i\hbar \frac{\partial \psi}{\partial t} = \left(-\frac{\hbar^2}{2m} \nabla^2
+V_{trap} + g N |\psi|^2 \right) \psi  \;,
\label{GP}
\end{equation}
where $g=4\pi\hbar^2 a/m$. The above equation is solved on a three-dimensional
grid using the Fast Fourier Transform split-operator method. We monitor the
diagonal part of the single-particle density matrix that is just the density of 
the classical field.

\begin{figure}[htb]
\resizebox{3.0in}{2.4in}
{\includegraphics{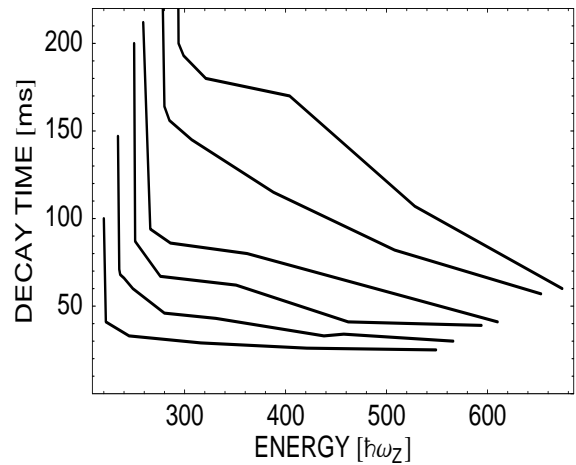}}
\caption{Decay time of doubly quantized vortex as a function of the energy per 
particle. The successive curves correspond to different values of the control 
parameter defined as $an_z$, where $a$ is the scattering length and $n_z$ is an 
axial density at the center of the trap. From top to bottom $an_z$ equals
$12.2$, $10.0$, $7.2$, $5.9$, $3.9$, and $2.3$, respectively.}
\label{decay}
\end{figure}

\begin{figure}[htb]
\resizebox{3.2in}{1.6in}
{\includegraphics{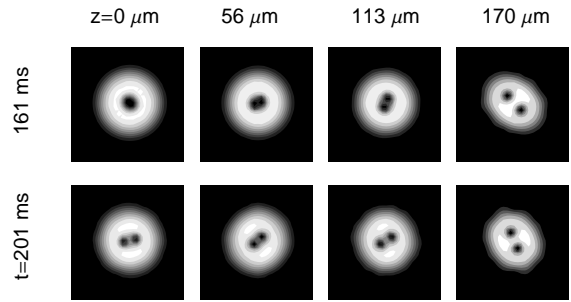}}
\caption{Density cuts at zero temperature and $an_z=10.0$ for different times showing 
that even at zero temperature the doubly quantized vortex eventually decays and two 
singly quantized vortices become disentangled along the whole condensate.}
\label{density}
\end{figure}

In Fig. \ref{decay} we plot the time needed to split the doubly quantized vortex
as a function of the energy pumped into the system while introducing the thermal
noise. For each curve ({\it i.e.}, the value of the interaction strength) the considered 
energies range from the zero temperature energy to an amount approximately twice larger. 
We have checked that this corresponds at most to $20\%$ condensate depletion. The basic 
observation is that the decay time is extremely sensitive to the presence of the noise. 
The rise of the energy by less than $1\%$ (in comparison with the zero temperature
energy) already results in a violant decrease of the decay time. For example, it drops 
from $200$\,ms to $87$\,ms for $an_z=5.9$ (the case of zero temperature energy per 
particle equal to $250\,\hbar\omega_z$). Rigorously speaking, 
at zero temperature and without any disturbances that might initialize the decay (like
the trap anisotropy or any deficiencies related to the process of phase imprinting)
the lifetime of the vortex should be infinite. Indeed, curves in Fig. \ref{decay}
show well visible asymptotes when approaching zero temperature energies (although
the top-lying end of each curve is a zero temperature decay time that is finite due
to the presence of numerical noise). Further increase of the amount of the thermal 
noise causes additional, although slower, decrease of the decay time. Fig. \ref{decay} 
shows that the decay time becomes comparable with the experimentally measured values 
(tens of milliseconds) only when the thermal noise at the appropriate level is included 
in the dynamics.

Even at zero temperature the vortex decays (see Fig. \ref{density}). The decay 
process sets in independently of the density as opposed to what is predicted  based 
on two-dimensional spectral analysis of the Bogoliubov eigenmodes. According to this 
analysis, the stability windows with respect to the interaction strength exist, the 
first one appearing approximately for \mbox{$3<an_z<12$}. They are defined by the 
requirement that all eigenfrequencies of all excitation modes are real. Within the 
stability window no decay of the doubly quantized vortex is expected. Outside the 
window, {\it i.e.}, when there is a complex eigenfrequency, the corresponding excitation 
mode grows exponentially in time resulting in the splitting of the vortex. Adding the 
third dimension changes the results dramatically as it is shown in Figs. \ref{decay} 
and \ref{density}. Splitting of the vortex in three-dimensional condensate need not 
be a simple and uniform process. In fact, it starts in the region where the local 
value of $an_z$ is in the two-dimensional instability window (see Ref. \cite{Mottonen}). 
So, the vortex splitting begins at the ends of the condensate (assuming $an_z < 12$) 
and propagates towards the center of the trap. In Fig. \ref{decay} we plot the lifetime 
of the doubly quantized vortex understood just as the time needed to split the vortex 
and disentangle it along the whole condensate. 

\begin{figure}[bth]
\resizebox{2.9in}{3.9in}
{\includegraphics{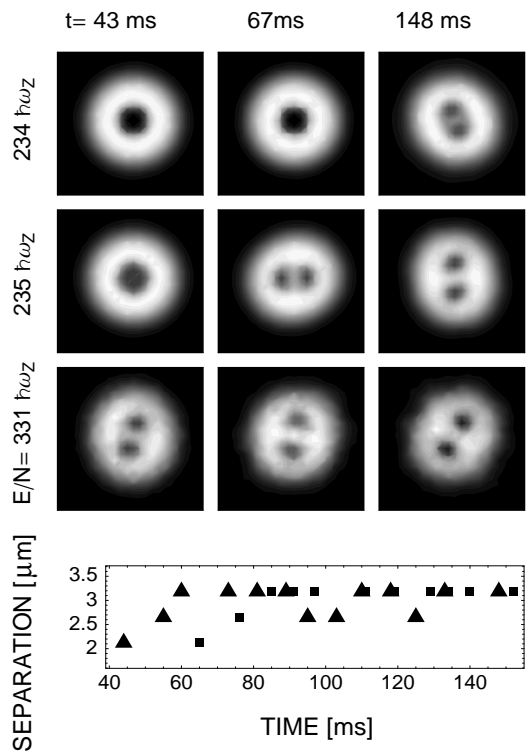}}
\caption{Tomographic images of the condensate that illustrate the dependence of
the decay time on the energy of the system. The interaction strength equals 
$an_z = 3.9$ and the energy per particle is $234$, $235$, and $331$ $\hbar \omega_z$
from top to bottom, respectively. The lowest panel shows the separation of two 
singly quantized vortices in the case when $E/N=235\, \hbar\omega_z$ (squares)
and $E/N=331\, \hbar\omega_z$ (triangles).}
\label{tomog}
\end{figure}

In Fig. \ref{tomog} we compare the decay time of a doubly quantized vortex for
different levels of the introduced thermal noise. The pictures are obtained according 
to the tomographic imaging technique as described in Ref. \cite{Ketterle2}. In this 
method, atoms within a $30$ $\mu$m thick central slice of the condensate are transfered 
to a different hyperfine state and then imaged by using another resonant laser pulse.
Following this prescription, each frame in Fig. \ref{tomog} shows the contour
plots of the density integrated axially within the slice of $30$ $\mu$m thickness
located at the center of the trap. The first two rows demonstrate how important is 
the thermal noise. Introducing the thermal noise on a very low level (the energy of the 
system is increased by less than $1\%$) already decreases the time needed to split 
the vortex by half. Rising the level of the thermal noise causes further lowering 
of the decay time what is necessary to get an agreement with the experiment.
In the bottom panel we plot the separation between two singly quantized vortices 
as a function of time. The case when the energy of the system is increased by 
approximately $50\%$ in comparison with the zero temperature energy is denoted by 
triangles. Although the population of the condensate is now about $94\%$, the 
distance between the two cores ($\sim 3\,\mu$m) almost does not change over the 
period of $90$\,ms. 


\begin{figure}[thb]
\resizebox{3.2in}{2.3in}
{\includegraphics{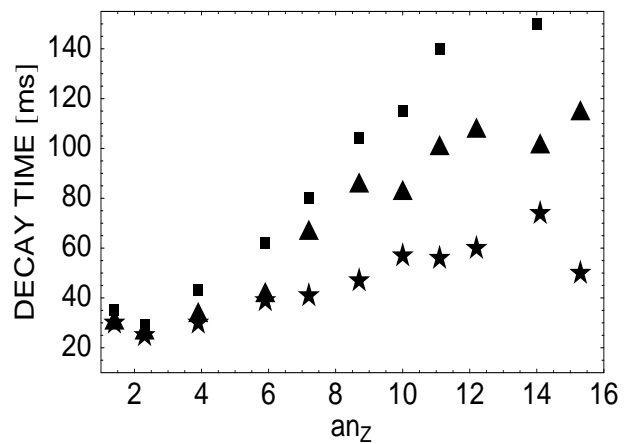}}
\caption{Decay time of doubly quantized vortex as a function of the density.
Three sets of points correspond to different values of the energy brought in
the system while introducing the thermal noise. From top to bottom the relative
increase of the energy equals $40\%$ (squares), $85\%$ (triangles), and $135\%$
(stars). The condensate depletion amounts to $6\%$, $13\%$, and $20\%$, 
respectively.}
\label{main}
\end{figure}

Finally, in Fig. \ref{main} we present the data regarding the time of the decay
of doubly quantized vortices which are intended to reproduce the main result of
the experimental work \cite{Ketterle2}. The authors of Ref. \cite{Ketterle2}
say only that the experiment was performed at the lowest possible temperature. 
Since the temperature of the system is determined based on the expansion of
the thermal cloud, this statement puts on, in fact, the constraint on the number of 
thermal atoms. To our knowledge the presence of less than $15\%$ of thermal atoms
can not be detected by using currently available experimental techniques. Therefore, 
in Fig. \ref{main} we plot three sets of data, each of them corresponding to different
level of thermal noise. More precisely, the relative increase of the energy of the 
system is constant for each data set. It is clear from Fig. \ref{main} that the decay
time gets larger when the interaction strength increases what remains in agreement
with the experiment. For larger density, however, the decay time is getting saturated.

\begin{figure}[thb]
\resizebox{3.0in}{2.0in}
{\includegraphics{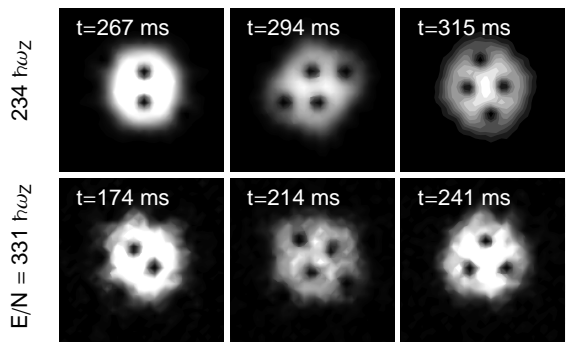}}
\caption{Density cuts along the radial plane $z=0$ at zero (the first row) and finite 
(the bottom row) temperatures for the interaction strength $an_z = 3.9$.}
\label{additional}
\end{figure}

To answer the question whether other sources of instability are able to reproduce
the experimental results of Ref. \cite{Ketterle2} we investigated the influence 
of the trap anisotropy (as high as $2.3\%$ according to \cite{private}) caused by
the gravitational sag as well as imprinting deficiencies related to the rapid change
of the magnetic field producing the vortex and resulting in radial squeeze and vertical
kick of the condensate \cite{Ketterle2,private}. We checked that the trap anisotropy
makes the lifetime of a doubly quantized vortex finite but still bigger than the
experimental values. For example, for $an_z=10.0$ the decay time equals $170\,$ms
whereas for $an_z=3.9$ it is $76\,$ms. Moreover, since the gravitational field do
not fluctuate the trap anisotropy can not explain the huge scatter of data in Fig. 3
of Ref. \cite{Ketterle2}. Shifting the whole condensate off the center of the trap
also does not help. It makes the total energy higher by increasing the potential
energy not the kinetic one. Therefore the number of uncondensed atoms remains constant
and the lifetime of the vortex equals that for unshifted condensate. Finally, we 
considered the dynamical phase imprinting instead of the topological one. In this case 
the energy pumped into the system goes to the kinetic energy and results in a production 
of uncondensed atoms. The decay time is comparable with the experimental values revealing 
in this way the basic role of the uncondensed atoms in the process of decaying of doubly 
quantized vortices.

Another interesting phenomenon is discovered while tracing the dynamics of the splitting
and subsequent evolution of two singly quantized vortices for longer times. It turns out 
that additional vortices enter the condensate. In Fig. \ref{additional} the first row 
corresponds to the case of zero temperature. Here, we observe that two singly quantized 
vortices (as numerically checked by using the interference technique as described in 
\cite{inter}) enter the condensate at the same time and settle into a lattice (in a rotating 
frame of reference) realizing the first scenario reported in Ref. \cite{Lobo}. At finite 
temperatures (the last row) the dynamics is reacher. First, two additional vortices 
simultaneously enter the condensate, later on, however, one vortex is lost and the system 
ends with the lattice consisting of three vortices.

In conclusion, we have addressed the issue concerning the lifetime of a doubly
quantized vortex raised by a recent experiment \cite{Ketterle2}. We show that the
decay of a doubly quantized vortex is driven rather by the thermal fluctuations than
other kinds of perturbation (quantum fluctuations since we need $10$ to $15\%$ of
thermal atoms to achieve the agreement with experiment \cite{Ketterle2} or the trap anisotropy). The thermal noise is the only reason, we find, able to decrease the time needed 
to split the vortex and make it comparable to the experimental values. As a result, the 
decay time is a monotonic function of the interaction strength. Our stressing of the role 
of thermal noise does not contradict the statement of the authors of Ref. \cite{Ketterle2} 
saying that the experiment was performed at the lowest possible temperature.

\acknowledgments 
We thank M. Gajda for helpful discussions.
We acknowledge support by the Polish Ministry of Scientific Research 
Grant Quantum Information and Quantum Engineering
No. PBZ-MIN-008/P03/2003.

\end{document}